\documentclass[aps,prl,reprint,twocolumn,superscriptaddress]{revtex4-2}
\usepackage{amssymb}
\usepackage{amsmath}
\usepackage{graphicx}
\usepackage{amsmath}
\usepackage{amssymb}
\usepackage{hyperref}
\hypersetup{
    colorlinks=true,
    linkcolor=blue,
    citecolor=blue,
    urlcolor=blue
}

\begin{document}

\title{Scaling of a Mutual-Information Distance in One-dimensional Quantum Spin Chains}
\author{Beau Leighton-Trudel}
\affiliation{Independent Researcher, Saint Petersburg, Florida, USA}
\date{November 27, 2025}

\begin{abstract}
We introduce a geometric scaling relation that characterizes the local scale behavior of correlations using the informational distance $d_E = K_0/\sqrt{I}$, where $I$ is the mutual information. We define a geometric conversion factor, $G \equiv \partial_r d_E$, which quantifies the local scale. We show that $G$ relates directly to $I$ via $G \propto I^{\kappa}$. For systems with power-law correlations $I(r) \sim r^{-X}$, the metric scaling exponent is $\kappa = 1/X - 1/2$. A key consequence is that the geometric scale $G$ is uniform (position-independent) if and only if $\kappa=0$, which occurs precisely at $X=2$. This identifies $X=2$ as the unique condition for a uniform and metric informational distance. We validate this relation using DMRG simulations of the 1D XXZ chain and exact results for the XX model. We demonstrate two falsifiable diagnostics: (i) $G(r)$ is flat in the bulk at criticality ($X\approx 2$) but varies strongly when gapped; (ii) a coordinate-agnostic slope test of $\log G$ versus $\log I$ at the XX benchmark ($X=2$) yields $\kappa \simeq 0$. This approach provides a coordinate-independent method for identifying scaling regimes that helps to reduce ambiguity from non-universal amplitudes and from the fitting choices in standard power-law analyses, and defines a simple post-processing pipeline that can be applied directly to numerical or experimental mutual-information data.
\end{abstract}

\maketitle

\section{Introduction}

Characterizing the scaling behavior of correlations in many-body systems is a central task in statistical mechanics ~\cite{Cardy1996, Sachdev2011}. While entanglement entropies and the quantum mutual information $I$ 
provide powerful measures of correlations in many-body systems~\cite{Amico2008,EisertCramerPlenio2010,Laflorencie2016}, extracting critical exponents ($X$) from coordinate-dependent fits ($I(r) \sim r^{-X}$) is often complicated by finite-size effects, boundary conditions~\cite{Affleck1986}, and finite-entanglement/finite-resolution effects in numerical fits~\cite{Tagliacozzo2008}. This motivates the search for robust diagnostics that can identify scaling regimes without relying on explicit coordinate dependence or non-universal amplitudes.

Recent work~\cite{LeightonTrudel2025} introduced a calibrated informational distance, $d_E = K_0/\sqrt{I}$, to probe the geometric structure of correlations. This definition is uniquely fixed by a Euclidean benchmark: $I(r) \propto r^{-2}$ maps to $d_E(r) \propto r$. For systems with power-law correlations -- as expected in 1D critical systems described by conformal field theory (CFT)~\cite{CalabreseCardy2004,CalabreseCardy2009}, this framework established a sharp criterion for metricity (satisfaction of the triangle inequality), which holds if and only if $0 < X \le 2$. While this addresses the global validity of the distance measure, it does not characterize its local scaling behavior—how the effective ``ruler'' changes with separation.

In this Letter, we investigate the local scale kinematics of this informational distance. We define a geometric conversion factor $G$ as the local rate of change of $d_E$:
\begin{equation}
    G(r) := \partial_r d_E(r).
    \label{eq:G_def}
\end{equation}
This factor quantifies the relationship between a coordinate increment $dr$ and the corresponding informational length increment $ds_E = G\,dr$.

We derive a direct relationship between this conversion factor and the mutual information, which we term the geometric scaling relation (GSR):
\begin{equation}
    G \propto I^\kappa,
    \label{eq:GSR}
\end{equation}

where $\kappa$ is the metric scaling exponent. For power-law mutual information $I(r) \sim r^{-X}$, this exponent is uniquely determined:
\begin{equation}
    \kappa = \frac{1}{X} - \frac{1}{2}.
    \label{eq:kappa}
\end{equation}

A crucial consequence of the GSR is that the geometric conversion factor $G$ is position-independent (constant) if and only if $\kappa=0$. This occurs precisely at the $X=2$ benchmark. This highlights the unique status of the $X=2$ regime: it is the only condition under which the informational distance is both metric (as established in~\cite{LeightonTrudel2025}) and possesses a uniform spatial scale.

Away from $X=2$, $G(r)$ varies with position, providing a fingerprint of the underlying phase. In the metric regime $0 < X < 2$, $\kappa > 0$, implying $G(r)$ decreases as correlations decay. In gapped phases, where $I(r)$ decays exponentially, $G(r)$ increases exponentially.

Practically, the GSR provides a coordinate-agnostic and amplitude-independent diagnostic. The relationship in Eq.~(\ref{eq:GSR}) implies that the slope of $\log G$ versus $\log I$ directly yields $\kappa$. This "slope test" bypasses the need for explicit $r$-dependent fits and filters out non-universal amplitudes, offering a robust method for identifying the $X=2$ regime ($\kappa=0$).

We validate this relation using exact analytical results for the 1D XX chain and high-precision DMRG simulations for the XXZ chain~\cite{FurusakiHikihara1998}. At the XX point (where $X=2$ is expected), $G(r)$ is demonstrably flat in the bulk, and the slope test yields $\kappa \approx 0$. In contrast, gapped phases show strong variation in $G(r)$. This provides a simple, falsifiable test that relies solely on intrinsic information quantities, bypassing the ambiguities of traditional coordinate fits. The scope of this work is strictly a kinematic analysis of static correlations, using mutual information defined in a fixed local basis and partition; all phase-identification statements are made within this operational choice.

\section{The Geometric Scaling Relation}

We now establish the central theoretical result connecting the geometric conversion factor ($G$) directly to the mutual information ($I$), independent of the coordinate $r$. This relationship arises from the interplay between the calibrated informational distance and the assumed power-law scaling of correlations.

We begin with the definitions established previously: the mutual information follows a power law $I(r) = C_I r^{-X}$ ($X>0, C_I>0$), and the informational distance is $d_E = K_0/\sqrt{I}$~\cite{LeightonTrudel2025}. Substituting the power law into the distance definition yields:
\begin{equation}
    d_E(r) = \frac{K_0}{\sqrt{C_I r^{-X}}} = K_0 C_I^{-1/2} r^{X/2}.
\end{equation}
The geometric conversion factor $G(r) = \partial_r d_E(r)$ is then obtained by differentiation:
\begin{equation}
    G(r) = \frac{X}{2} K_0 C_I^{-1/2} r^{X/2 - 1}.
    \label{eq:G_vs_r}
\end{equation}

To derive the geometric scaling relation (GSR), we eliminate the explicit dependence on the coordinate $r$. We invert the mutual information power law to express $r$ in terms of $I(r)$: $r = C_I^{1/X} [I(r)]^{-1/X}$. Substituting this expression for $r$ back into Eq.~(\ref{eq:G_vs_r}) yields:
\begin{align}
    G(r) &= \frac{X}{2} K_0 C_I^{-1/2} \left( C_I^{1/X} [I(r)]^{-1/X} \right)^{X/2 - 1} \\
    &= \frac{X}{2} K_0 C_I^{-1/2} C_I^{1/2 - 1/X} [I(r)]^{-1/2 + 1/X} \\
    &= A \cdot [I(r)]^{\kappa}.
    \label{eq:GSR_main}
\end{align}
This confirms the GSR, $G \propto I^\kappa$, where the metric scaling exponent $\kappa$ is
\begin{equation}
    \kappa = \frac{1}{X} - \frac{1}{2},
    \label{eq:kappa_main}
\end{equation}
and the amplitude $A = \frac{X}{2} K_0 C_I^{-1/X}$ is constant for a given scaling regime.

\subsection{Implications and the Coordinate-Agnostic Slope Test}

The geometric scaling relation reveals a sharp connection between the information exponent $X$ and the spatial uniformity of the informational distance scale.

\textbf{The Uniform Scale Condition ($X=2$).} The geometric conversion factor $G(r)$ is constant (position-independent) if and only if the metric scaling exponent $\kappa=0$. According to Eq.~(\ref{eq:kappa_main}), this occurs precisely at the $X=2$ benchmark used for calibration. This condition uniquely identifies the regime where the conversion between coordinate distance and informational distance is uniform across the system.

\textbf{Behavior Across Phases.} Away from $X=2$, $G(r)$ varies with position. Within the metric window $0 < X < 2$~\cite{LeightonTrudel2025}, the exponent $\kappa$ is positive. Since $I(r)$ decreases with $r$, $G(r)$ also decreases. (This can also be seen directly in Eq.~(\ref{eq:G_vs_r}), where the exponent $X/2-1$ is negative.) This signifies a scale that contracts at larger distances.

In gapped phases, mutual information decays exponentially, $I(r) \sim e^{-r/\xi}$~\cite{WolfVerstraeteHastingsCirac2008}. The informational distance grows as $d_E(r) \sim e^{r/(2\xi)}$. Consequently,
\begin{equation}
    G(r) = \partial_r d_E(r) \sim \frac{1}{2\xi} e^{r/(2\xi)}.
\end{equation}
Here, $G(r)$ is strongly non-constant, growing exponentially with separation. The flatness of $G(r)$ thus provides a clear diagnostic distinction between the $X=2$ critical regime and gapped phases.

\textbf{The Coordinate-Agnostic Slope Test.} The GSR provides a practical, falsifiable diagnostic for identifying scaling regimes. Taking the logarithm of Eq.~(\ref{eq:GSR_main}) yields a linear relationship:
\begin{equation}
    \log G = \kappa \log I + \log A.
    \label{eq:SlopeTest}
\end{equation}
The metric scaling exponent $\kappa$ is simply the slope of $\log G$ versus $\log I$. Crucially, this test is independent of the choice of origin and units for the coordinate $r$. Furthermore, the slope $\kappa$ is independent of the non-universal amplitudes $K_0$ and $C_I$, which only affect the intercept. In regimes with a power-law tail, the standard information exponent is recovered from $\kappa$ via $X = 1/(\kappa + 1/2)$, so the slope test provides an alternative, coordinate-independent route to exponent extraction. Measuring $\kappa=0$ in particular gives direct evidence of the $X=2$ scaling regime.

\section{Numerical Validation}

We validate the geometric scaling relation (GSR) and demonstrate its diagnostic utility using two complementary approaches in 1D spin chains. In both cases we follow the same simple analysis pipeline: (i) given a mutual-information profile $I(r)$, we form the informational distance $d_E(r) = 1/\sqrt{I(r)}$ (up to the overall scale $K_0$); (ii) we differentiate to obtain the geometric conversion factor $G(r) = \partial_r d_E(r)$; and (iii) we either inspect the flatness of $G(r)$ in a bulk window or fit the slope of $\log G$ versus $\log I$ to extract $\kappa$. 

First, we use Density Matrix Renormalization Group (DMRG) simulations of the XXZ model to analyze the position dependence of the geometric conversion factor $G(r)$. Second, we perform the coordinate-agnostic slope test on the exact solution of the XX model, providing a high-precision benchmark for the $X=2$ regime.

\textbf{Test 1: Position Dependence of $G(r)$ in the XXZ Chain.}
We investigate the prediction that $G(r)$ is constant when $X\approx 2$. We study the 1D spin-1/2 XXZ Hamiltonian,
\begin{equation}
    H = \sum_{i} \left( S^x_i S^x_{i+1} + S^y_i S^y_{i+1} + \Delta S^z_i S^z_{i+1} \right),
\end{equation}
on a chain of $L=96$ sites with open boundary conditions (OBC); see Ref.~\cite{FurusakiHikihara1998} and ~\cite{Giamarchi2004} for background on this model. We use DMRG~\cite{White1992, Schollwoeck2011} (maximum bond dimension $D=128$) to obtain the ground state in the critical ($\Delta=1.0$) and gapped antiferromagnetic ($\Delta=2.0$) phases.

To probe the bulk behavior, we measure the mutual information $I(r)$ between the central site and a site at separation $r$. We compute the informational distance $d_E(r) = 1/\sqrt{I(r)}$ (setting $K_0=1$ as the absolute scale does not affect the scaling exponent).

\begin{figure}[t]
    \centering
    \includegraphics[width=\columnwidth]{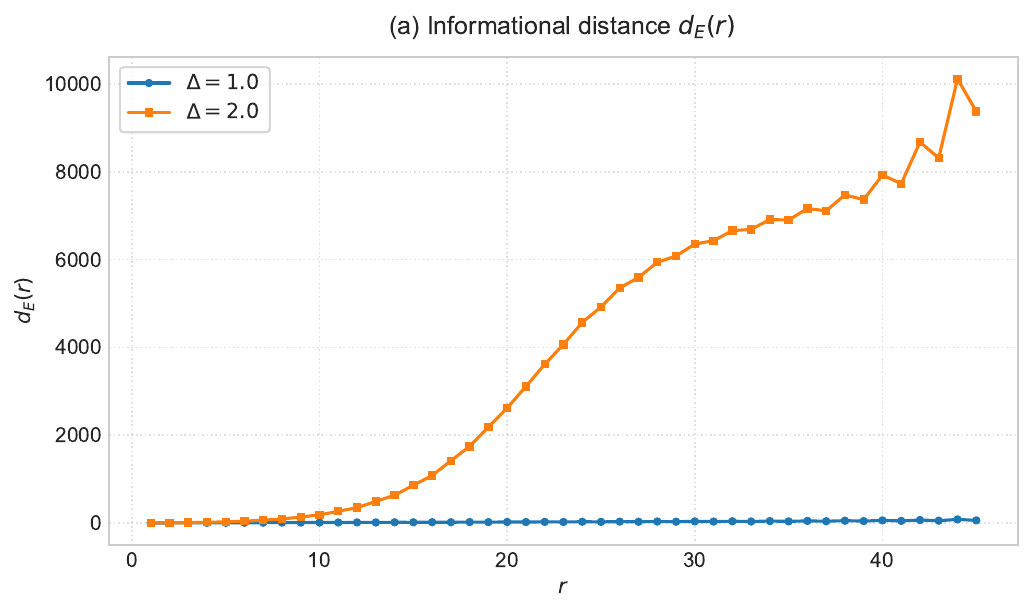}
    \includegraphics[width=\columnwidth]{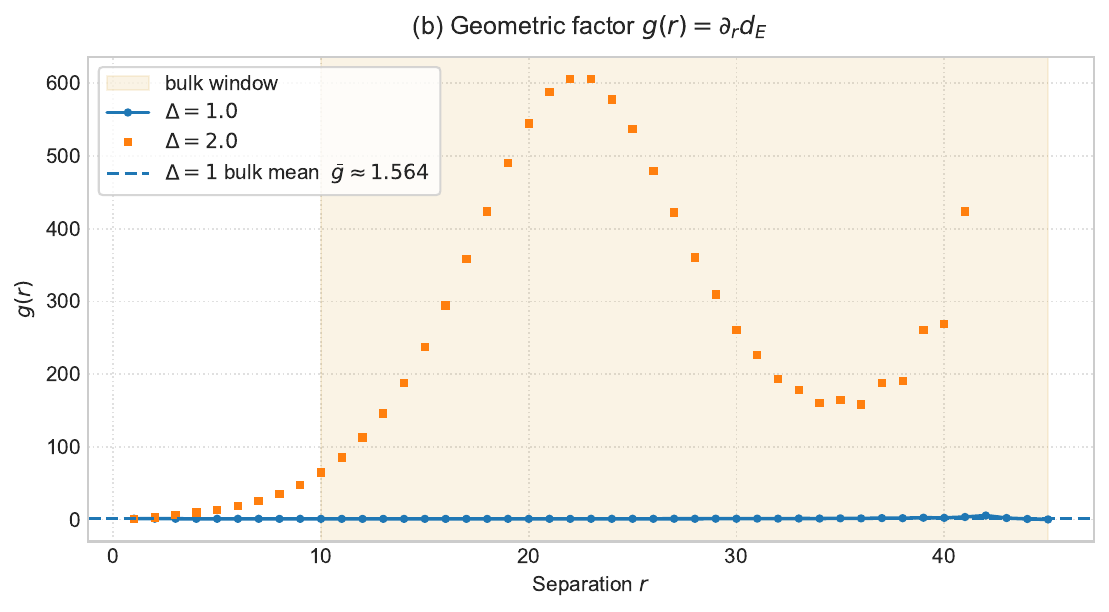}
   \caption{\textbf{Informational distance and geometric conversion factor in the XXZ chain.} (DMRG, $L=96$, OBC, $D=128$). (a) Informational distance $d_E(r)$ computed from center-pair mutual information. (b) Geometric conversion factor $G(r) = \partial_r d_E$. In the critical phase ($\Delta=1$), $G(r)$ is flat in the bulk window ($r\ge 10$, shaded), indicating a uniform scale ($\kappa \approx 0$). The dashed line shows the bulk mean $\bar{G} \approx 1.564$. In the gapped phase ($\Delta=2$), $G(r)$ grows rapidly and varies strongly with position, then turns over at the largest separations where $I(r)$ is at the numerical noise floor ($I(r) \lesssim 10^{-8}$); in this regime the derivative is not quantitatively reliable, but the strong non-uniformity of $G(r)$ relative to the critical case is already evident at intermediate $r$.}
    \label{fig:xxz_static}
\end{figure}

Figure~\ref{fig:xxz_static}(a) shows $d_E(r)$. At criticality ($\Delta=1.0$), $d_E(r)$ grows approximately linearly, consistent with $X\approx 2$ scaling. In the gapped phase ($\Delta=2.0$), $d_E(r)$ grows rapidly, consistent with exponential decay of $I(r)$.

To obtain the geometric conversion factor $G(r) = \partial_r d_E(r)$, we employ a careful numerical differentiation scheme. We first apply a Savitzky-Golay filter (window 7, polynomial order 2) to $d_E(r)$ to mitigate numerical noise from the DMRG data, and then use central finite differences on the uniform $r$-grid.

Figure~\ref{fig:xxz_static}(b) presents $G(r)$, revealing a stark contrast. We analyze the data within a bulk window ($r \ge 10$, shaded region) to minimize boundary effects. At criticality, $G(r)$ is demonstrably flat in the bulk. This flatness signifies a uniform geometric scale, consistent with the GSR prediction that $\kappa \approx 0$ leads to constant $G$. This is quantified by a small coefficient of variation (CoV = $\sigma/\mu \approx 0.0054$).

In the gapped phase, $G(r)$ varies strongly with position: for $\Delta=2.0$ the rapid increase is clearly visible for $10 \lesssim r \lesssim 25$. At larger separations the mutual information has dropped to the numerical floor of the DMRG calculation ($I(r) \sim 10^{-8}$), so $d_E(r)$ effectively saturates and the Savitzky--Golay/finite-difference procedure underestimates $\partial_r d_E$, producing the apparent downturn of $G(r)$ at the largest $r$. We therefore only plot $G(r)$ where the underlying $I(r) > 10^{-8}$ and interpret this tail as a finite-precision artifact; our diagnostic relies on the strong non-flatness of $G(r)$ in the intermediate-$r$ window compared to the nearly constant critical curve.

\textbf{Test 2: The Coordinate-Agnostic Slope Test.}
We now apply the coordinate-agnostic slope test (Eq.~(\ref{eq:SlopeTest})) to quantitatively verify the GSR. We utilize the 1D XX chain ($\Delta=0$) in the thermodynamic limit, which is exactly solveable ~\cite{LiebSchultzMattis1961} and provides an exact benchmark for the $X=2$ regime.

The mutual information is calculated analytically from the known ground-state correlators at half-filling, $C(r) = \sin(\pi r/2)/(\pi r)$ ~\cite{PeschelEisler2009}. We restrict the analysis to odd separations $r$ (up to $r=401$), as $I(r)=0$ for even $r$. We compute $I(r)$, $d_E(r)$, and $G(r)$ using central differences on the uniform odd-$r$ grid ($\Delta r=2$). As the data is exact, no smoothing is applied.

We then perform a linear fit of $\log G$ versus $\log I$ over a bulk window $r \ge 15$ ($n=194$ points). According to the GSR, the slope must be $\kappa = 1/X - 1/2$. For $X=2$, we predict $\kappa=0$.

\begin{figure}[t]
    \centering
    \includegraphics[width=\columnwidth]{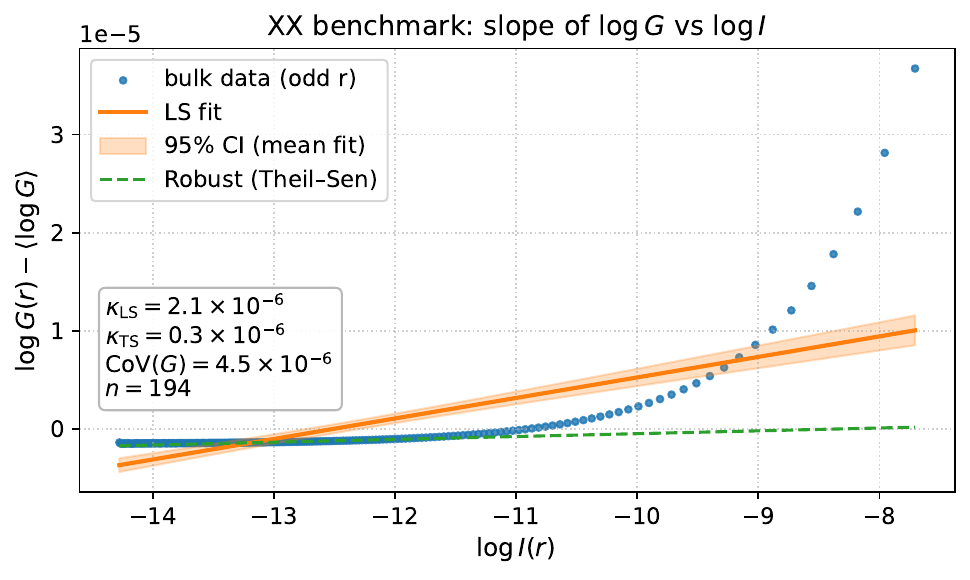}
    \caption{\textbf{Coordinate-agnostic slope test at the $X=2$ benchmark.} Exact results for the XX chain ground state (thermodynamic limit). The plot shows $\log G(r) - \langle \log G \rangle$ versus $\log I(r)$ for bulk data (odd $r \ge 15$). The slope $\kappa$ is extracted using least-squares (LS, orange line with 95\% CI band) and robust Theil-Sen (TS, green dashed) estimators. Both yield $\kappa \approx 0$ (inset), confirming the GSR prediction $\kappa = 0$ at $X=2$. The extremely small CoV($G$) quantifies the uniformity of the geometric scale.}
    \label{fig:xx_slope_test}
\end{figure}

The results are shown in Figure~\ref{fig:xx_slope_test}. The slope is vanishingly small. A standard least-squares (LS) fit yields $\kappa_{\text{LS}} = 2.1 \times 10^{-6}$. A robust Theil-Sen (TS) estimator confirms this, yielding $\kappa_{\text{TS}} = 0.3 \times 10^{-6}$. Both estimates are consistent with zero within the 95\% confidence interval. Furthermore, the uniformity of the scale $G$ is confirmed by its extremely small coefficient of variation (CoV($G$) $= 4.5 \times 10^{-6}$).

This result provides strong quantitative validation of the geometric scaling relation. It demonstrates that the slope test accurately identifies the $X=2$ regime ($\kappa=0$) using a method independent of coordinates and non-universal amplitudes.

\section{Interpretation and Scope}
\label{sec:scope}

The geometric scaling relation (GSR) provides a kinematic framework for analyzing the scale dependence of correlations directly from static mutual information. The analysis presented here is strictly dimensionless and focuses on the behavior of the global scaling function $d_E(r)$ and its derivative $G(r)$.

This approach is designed as a global diagnostic. We do not interpret $d_E$ as a fundamental real-space metric or as a proposal for new spatial geometry; throughout, $d_E$ and $G$ are used solely as diagnostic functionals of the mutual information. We emphasize the distinction established in~\cite{LeightonTrudel2025}: while $d_E$ constitutes a valid metric only within the window $0<X\le 2$, the GSR remains applicable as a diagnostic of scale behavior even in regimes where $d_E$ is not globally metric, such as in gapped phases or when $X>2$. In these cases, $G(r)$ and $\kappa$ still quantify the non-uniformity of the informational distance scale.

A sharp distinction is maintained between this global analysis and a local Riemannian geometry. We do not identify the global distance $d_E$ with a geodesic length derived from a local metric tensor, except in the specific case of the homogeneous, isotropic $X=2$ benchmark where the scale is uniform. Outside this benchmark, $d_E$ is treated strictly as a global diagnostic functional of $I$. The definition of local geometric quantities, such as angles or curvature, requires a separate analysis.

The diagnostic is operational, acknowledging that mutual information depends on the choice of basis and system partition. The GSR thus probes the geometry of scale relative to a specific definition of observables. (Robustness across different choices is assessed in the SM.) Furthermore, behavior under local coarse-graining is constrained by the Data Processing Inequality~\cite{CoverThomas2006}: coarse-graining cannot increase $I$, and thus cannot decrease $d_E$.

The coordinate-agnostic slope test utilizes the intrinsic relationship between $I$ and $G$. The resulting fit ($\log G=\kappa\log I+\mathrm{const}$) is independent of non-universal amplitudes and invariant under linear rescaling or shifts of the coordinate $r$. It is not, however, invariant under arbitrary nonlinear reparametrizations $s=\phi(r)$, which inherently modify the definition of $G=\partial_r d_E$.

While the main text addresses the isotropic 1D case, the framework generalizes. The SM details the direction-resolved GSR for anisotropic systems, $G(\hat{n}) \propto I(\hat{n})^{\kappa(\hat{n})}$, and the analysis of multiplicative logarithmic corrections relevant to specific critical points.

\section{Discussion and Outlook}

We have introduced the geometric scaling relation (GSR), a direct relationship between the local scale of informational distance and mutual information. By defining the geometric conversion factor $G = \partial_r d_E$, we derived the relation $G \propto I^\kappa$. For systems with power-law correlations $I(r) \sim r^{-X}$, the metric scaling exponent is uniquely determined as $\kappa = 1/X - 1/2$.

The central insight of this kinematic framework is the sharp criterion for a uniform geometric scale: $G$ is position-independent if and only if $\kappa=0$, which occurs precisely at $X=2$. This elevates the significance of the $X=2$ benchmark, identifying it as the unique condition where the informational distance is both globally metric~\cite{LeightonTrudel2025} and locally uniform.

Practically, the GSR offers a robust alternative to traditional coordinate-dependent fits. The coordinate-agnostic slope test—measuring $\kappa$ via the slope of $\log G$ versus $\log I$—filters out non-universal amplitudes and reduces common ambiguities related to boundary effects and window selection. As demonstrated in the XX and XXZ models, this provides a falsifiable, quantitative diagnostic for identifying the $X=2$ scaling regime and characterizing scale behavior across phases.

\vspace{0.5em}
\noindent\textbf{Outlook.} The geometric scaling relation provides a versatile tool for analyzing correlation structures in many-body systems, opening avenues for further investigation:

(i) \textbf{Characterizing Diverse Phases.} The analysis pipeline is amplitude-independent and readily applicable. A natural next step is to apply this diagnostic to a broader range of models, including those in higher dimensions and interacting systems, to map the prevalence of the $X=2$ regime versus other critical or gapped behaviors.

(ii) \textbf{Anisotropy and Complexity.} Utilizing the direction-resolved exponent $\kappa(\hat{n})$ (detailed in the SM) allows for the characterization of anisotropic scaling. Furthermore, developing windowed estimators for $\kappa$ and flatness statistics for $G$ will enable the tracking of crossovers between different scaling regimes in disordered or complex systems.

(iii) \textbf{Experimental Application.} The framework can be directly applied to experimental datasets where mutual information, or suitable proxies, are accessible, offering a coordinate-independent method for analyzing scaling behavior.

(iv) \textbf{Dynamical Constraints.} Future work may investigate how the static geometric scaling identified here constrains the dynamics of information propagation.

(v) \textbf{Drop-in analysis for existing data.} Because the diagnostic only requires a mutual-information profile $I(r)$, it can be applied retrospectively to existing numerical datasets as a lightweight post-processing step, providing an additional, amplitude-independent check on phase identification and exponent estimates.

In summary, the geometric scaling relation provides a compact, reproducible method to read the scale behavior of quantum matter directly from mutual information, offering a distinct and robust diagnostic for statistical mechanics.

\bibliographystyle{apsrev4-2}
\bibliography{references}

\onecolumngrid
\begingroup
  \setlength{\parindent}{0pt}
  \setlength{\parskip}{0.8\baselineskip}

\section*{Acknowledgments}

We acknowledge the open-source developer community-particularly the teams behind \texttt{NumPy}, \texttt{SciPy}, \texttt{Quimb}, \texttt{Matplotlib}, and related libraries-which made these simulations and analyses possible.

\section{Code and Data Availability}

All source code used to generate figures and reproduce simulations is available at Zenodo: https://doi.org/10.5281/zenodo.17727059.
\end{document}